\documentclass[a4paper,11pt]{article}
\pdfoutput=1
\usepackage{graphicx}
\usepackage{epsf,amsmath,bbold,amsfonts,stmaryrd}

\usepackage[utf8]{inputenc}
\usepackage{mathrsfs}
\usepackage{appendix}
\usepackage{amssymb}
\usepackage{float}
\usepackage{color}
\usepackage{cite}
\usepackage{hyperref}
\hypersetup{pageanchor=false}
\usepackage{indentfirst}
\usepackage{url}
\usepackage{float}
\usepackage{caption}
\usepackage[numbers,square,comma,sort&compress,merge]{natbib}

\newcommand{\be}{\begin{equation}}
\newcommand{\ee}{\end{equation}}
\newcommand{\bea}{\setlength\arraycolsep{2pt} \begin{eqnarray}}
\newcommand{\eea}{\end{eqnarray}}
\newcommand{\nn}{\nonumber}

\def\ft#1#2{{\textstyle{\frac{\scriptstyle #1}{\scriptstyle #2} } }}
\def\fft#1#2{{\frac{#1}{#2}}}

\def\0{{\sst{(0)}}}
\def\1{{\sst{(1)}}}
\def\2{{\sst{(2)}}}
\def\3{{\sst{(3)}}}
\def\4{{\sst{(4)}}}
\def\5{{\sst{(5)}}}
\def\6{{\sst{(6)}}}
\def\7{{\sst{(7)}}}
\def\8{{\sst{(8)}}}
\def\sst#1{{\scriptscriptstyle #1}}

\begin{document}
\title{\textbf{Influence of Quantum Correction on the Black Hole Shadows, Photon Rings and Lensing Rings }}
\author{Jun Peng$^{1}$,
Minyong Guo$^{2}$ and Xing-Hui Feng$^{3*}$}
\date{}
\maketitle

\vspace{-10mm}

\begin{center}
{\it$^1$Van Swinderen Institute, University of Groningen, 9747 AG Groningen, The Netherlands\\\vspace{1mm}

$^2$Center for High Energy Physics, Peking University,
 Beijing 100871, China\\\vspace{1mm}

$^3$Center for Joint Quantum Studies and Department of Physics,
School of Science, Tianjin University, Tianjin 300350, China\\\vspace{1mm}
}
\end{center}

\vspace{8mm}

\begin{abstract}
  We calculate the photon sphere $r_{ph}$ and critical curve $b_c$ for the quantum corrected Schwarzschild black hole, finding that they violate an universal inequalities proved for asymptotically flat black holes which satisfy null energy condition in the framework of Einstein gravity. This violation seem to be a common phenomenon when considering quantum modification of Einstein gravity. Furthermore, we study the shadows, lensing rings and photon rings in the quantum corrected Schwarzschild black hole. The violation leads to a larger bright lensing ring in the observational appearance of thin disk emission near the black hole compared with the classical Schwarzschild black hole. Our analysis may provide a observational evidence for quantum effect of general relativity.
\end{abstract}

\vfill{\footnotesize jun.peng@rug.nl,\,\, minyongguo@pku.edu.cn,\,\, xhfeng@tju.edu.cn,\\$~~~~~~*$ Corresponding author.}

\maketitle

\newpage
\section{Introduction}
In classical general relativity, the problem of singularities has remained in suspense since it was pointed out by Penrose and Hawking \cite{Penrose:1964wq,Hawking:1969sw}. The reason why physicists hate singularities is mainly that the existence of singularities will lead to incomplete geodesics so that the causality is destroyed. However, the strong brief in the inevitability of causality persuade researches the classical general relativity fails and quantum effect must be considered around the singularity. Along this line, some interesting works have been made in the past years \cite{Stelle:1976gc,Stelle:1977ry,Biswas:2011ar,Biswas:2013cha,Biswas:2016egy}.

Quantum corrected black holes seem to be different from classical black holes in terms of the photon sphere and shadows since the quantum effect is significant in the strong gravitational region. For spherically symmetric black holes in classical general relativity, a lot of evidences shows that the radius of the horizon, shadow, and outermost photon sphere obey an universal conjecture \cite{Lu:2019zxb,Feng:2019zzn}
\be
\frac{3}{2}r_0\le r_{ph}\le\frac{b_c}{\sqrt{3}}\le3M\label{inequality}
\ee
where $M$ is the mass of black hole. Later, these inequalities were proved for asymptotically flat black holes which satisfy null energy condition in the framework of Einstein gravity in \cite{Yang:2019zcn}. However, by considering the spherically symmetric black hole in regularized $4D$ Einstein-Gauss-Bonnet (EGB) gravity, the authors found the relations are reversed when the GB coupling constant is negative \cite{Guo:2020zmf,Zeng:2020dco}, viz.
\be
\frac{3}{2}r_0\ge r_{ph}\ge\frac{b_c}{\sqrt{3}}\ge3M\label{rinequality}
\ee
In addition, the regularized $4D$ EGB black hole can also be obtained by considering conformal anomaly of Einstein gravity \cite{Cai:2009ua}. Moreover, the violation of inequalities was also found in \cite{Hennigar:2018hza} for Einsteinian cubic gravity and in \cite{Khodabakhshi:2020hny} for Einsteinian quartic gravity. It's worth noting that these theories are all well posed higher derivative gravities which would natural arise for ultraviolet complete quantum gravity such as string theory. Hence, one may guess that quantum corrections may cause the inequality to flip. In general an exact black hole solution is difficult to obtain when considering quantum  modification. In the first part of this paper, we will consider a simple quantum corrected Schwarzschild black hole obtained by Kazakov and Solodukhin \cite{Kazakov:1993ha} to test this conjecture.

Moreover, as we know shadow is a kind of darker area which is formed when an opaque body imprison the transmit of light. By definition, the shape of the shadow would depend on the location and intensity of the light source. The same is true for black hole shadows. For the standard usage of the term "shadow", the light source is thought to be distributed throughout the entire space desultorily, including from behind the observer. But if a particular light source (like accretion disk) near the black hole is so bright that the natural light sources can be ignored, the appearance of a black hole cannot described by the standard "shadow". To avoid ambiguity, we will call the standard "shadow"  the "critical curve" as suggested in \cite{Gralla:2019xty}. For example, the shadow we mentioned in the inequalities in the previous paragraph is the critical curve. The observed size of the central dark area is mainly governed by the emission profile and gravitational redshift, so there is more physical information in the so called photon ring and lensing ring \cite{Gralla:2019xty,Himwich:2020msm,Gralla:2020nwp,Gralla:2020yvo,Zeng:2020vsj}.

To gain a more complete picture of the effects of quantum corrections on the appearance of a black hole \footnote{In addition,
the optical appearance of a star orbiting a black hole is also  an interesting topic to study, see examples in \cite{Cunningham:1972, Cunningham:1973, Porfyriadis:2016gwb, Gralla:2017ufe, Guo:2018kis, Long:2018tij, Yan:2019etp, Guo:2019lur, Guo:2019pte, Li:2020val}. Furthermore, the shadow of a black hole observed by other physical observers also has people's attention, some interesting results can be found in \cite{Chang:2020miq, Perlick:2018iye, Bisnovatyi-Kogan:2018vxl, Li:2020drn}.}, we will give a full analysis on the shadows, photon rings and lensing rings taking the Kazakov-Solodukhin (KS) black hole as a case. In section 2, we obtain the radius of photon sphere and shadow of KS black hole and verify the inverse inequalities \eqref{rinequality}. In section 3, we analyze the light bending in KS black hole. In section 4, we calculate the innermost stable circular orbit (ISCO) in KS black hole. In section 5, we consider thin disk emission near the black hole and investigate the observational appearance with three typical emission profiles. We conclude in section 6.

\section{Null geodesic in quantum corrected Schwarzschild black hole}
The deformation of the Schwarzschild black hole due to spherically symmetric quantum fluctuations was discussed by Kazakov and Solodukhin \cite{Kazakov:1993ha}. The Kazakov-Solodukhin metric takes the form
\be
ds^2=-f(r)dt^2+\frac{dr^2}{f(r)}+r^2d\Omega^2_2\label{metric}
\ee
with
\be
f(r)=\frac{\sqrt{r^2-a^2}}{r}-\frac{2M}{r}\label{quantumbh}
\ee
where $a$ is the deformation parameter. The horizon is located at
\be
r_0=\sqrt{4M^2+a^2}
\ee
It should be pointed out that though the metric is finite at $r=a$, the curvature is divergent on this two-dimensional sphere.

The orbit of one particle traveling in curved spacetime is described by geodesic equations which can be expressed as the Euler-Lagrange equation
\be
\fft{d}{d\tau}\Big(\fft{\partial{\cal L}}{\partial \dot x^\mu}\Big)=\fft{\partial{\cal L}}{\partial x^\mu}
\ee
where $\tau$ is the proper time, $\dot x^\mu$ is the four-velocity of particle and ${\cal L}$ is the Lagrangian. For the line element \eqref{metric}, the Lagrangian is
\be
{\cal L}=\fft12g_{\mu\nu}\dot x^\mu\dot x^\nu=\fft12\Big(-f\dot t^2+\fft{\dot r^2}{f}+r^2\dot\theta^2+r^2\sin^2\theta\dot\phi^2\Big)
\ee
For static and spherically symmetric metric, we can always restrict the particle moving on the equatorial plane, i.e. $\theta=\ft\pi2$ and $\dot\theta=0$. In addition, the Lagrangian ${\cal L}$ doesn't involve coordinates $t$ and $\phi$ explicitly, so we have two conserved quantities
\be
E=-\fft{\partial{\cal L}}{\partial\dot t}=f\dot t,\quad L=\fft{\partial{\cal L}}{\partial\dot\phi}=r^2\dot\phi^2
\ee
as the respective energy and angular momentum of one particle travelling around the black hole.
For null geodesic ${\cal L}=0$, then we can obtain the orbit equation
\be
\Big(\frac{dr}{d\phi}\Big)^2=V_{eff}
\ee
with the effective potential given by
\be
V_{eff}=r^4\Big(\frac{1}{b^2}-\frac{f(r)}{r^2}\Big)
\ee
where $b=\ft LE$ is called the impact parameter. The circular orbit corresponds to
\be
V_{eff}=0,\quad V'_{eff}=0
\ee
Because of the spherical symmetry, all circular orbits form a closed surface which is the so called photon sphere whose radius is determined by
\be
\Big(\frac{f(r_{ph})}{r_{ph}^2}\Big)'=0
\ee
and the corresponding impact parameter $b_c$ gives the standard shadow radius
\be
b_c=\frac{r_{ph}}{\sqrt{f(r_{ph})}}
\ee
For quantum corrected black hole $\eqref{quantumbh}$, the photon sphere and shadow radius are respectively
\bea
r_{ph}&=&\sqrt{\frac{3(3+x^2+\sqrt{9+2x^2})}{2}}M\\
b_c&=&\sqrt{\frac{\sqrt{27(3+x^2+\sqrt{9+2x^2})^3}}{2\sqrt{9+x^2+3\sqrt{9+2x^2}}-4\sqrt2}}M
\eea
where we introduce a dimensionless parameter $x=a/M$. It is easy to show that for quantum corrected black hole, we have the reversed inequalities
\be
\fft32r_0 \ge r_{ph} \ge \frac{b_c}{\sqrt3} \ge 3M
\ee
as we have guessed in the introduction. These facts support the idea that the conjecture proposed in \cite{Lu:2019zxb,Feng:2019zzn} will be destroyed for quantum corrected black holes. And the underlying reason may be that the quantum corrected black holes usually don't respect the null energy condition.

\section{Light bending: diret, lensed and photon ring}\label{bending}
\begin{figure}[t!]
\begin{center}
\includegraphics[scale=0.5]{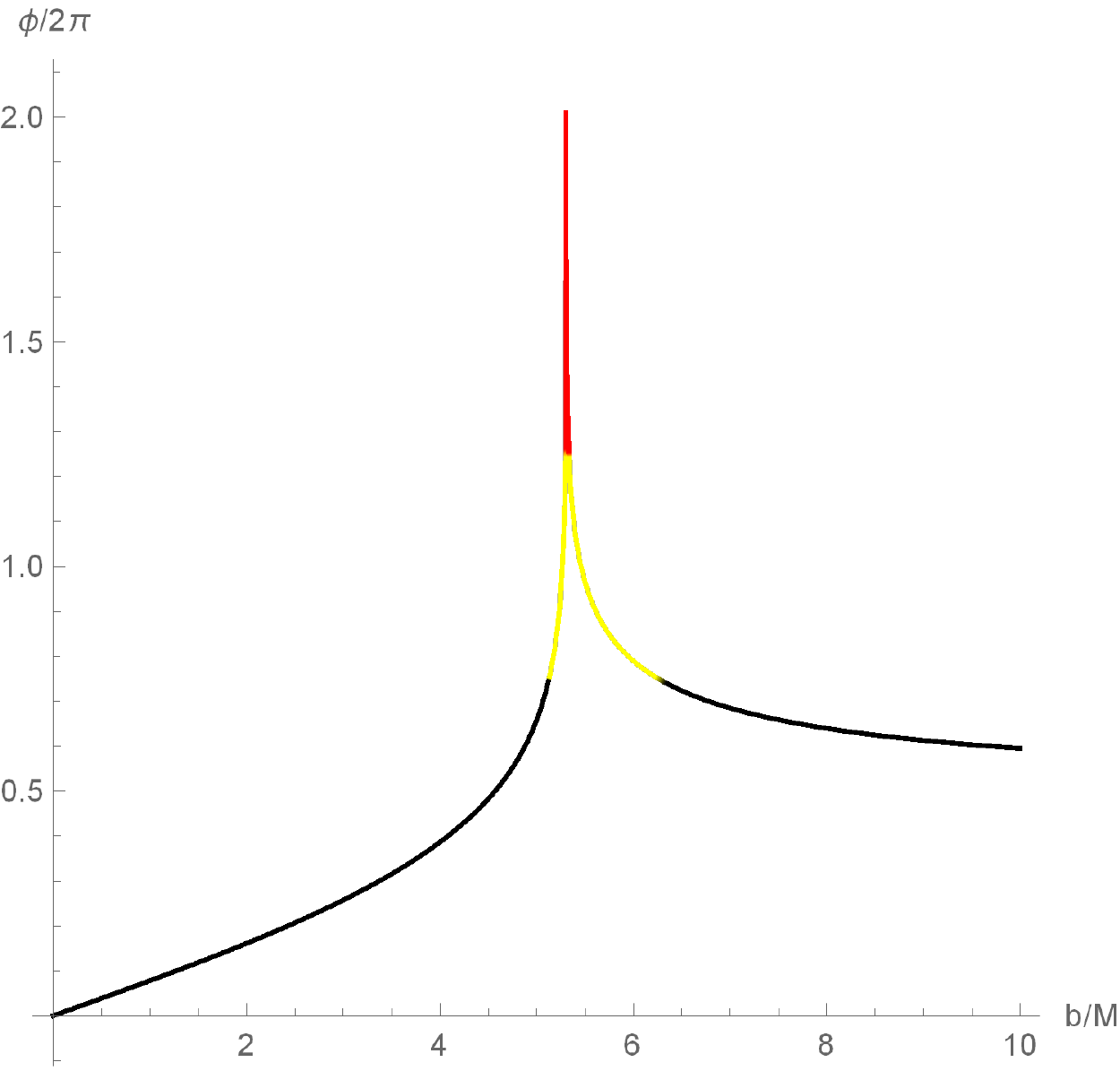}\qquad\includegraphics[scale=0.5]{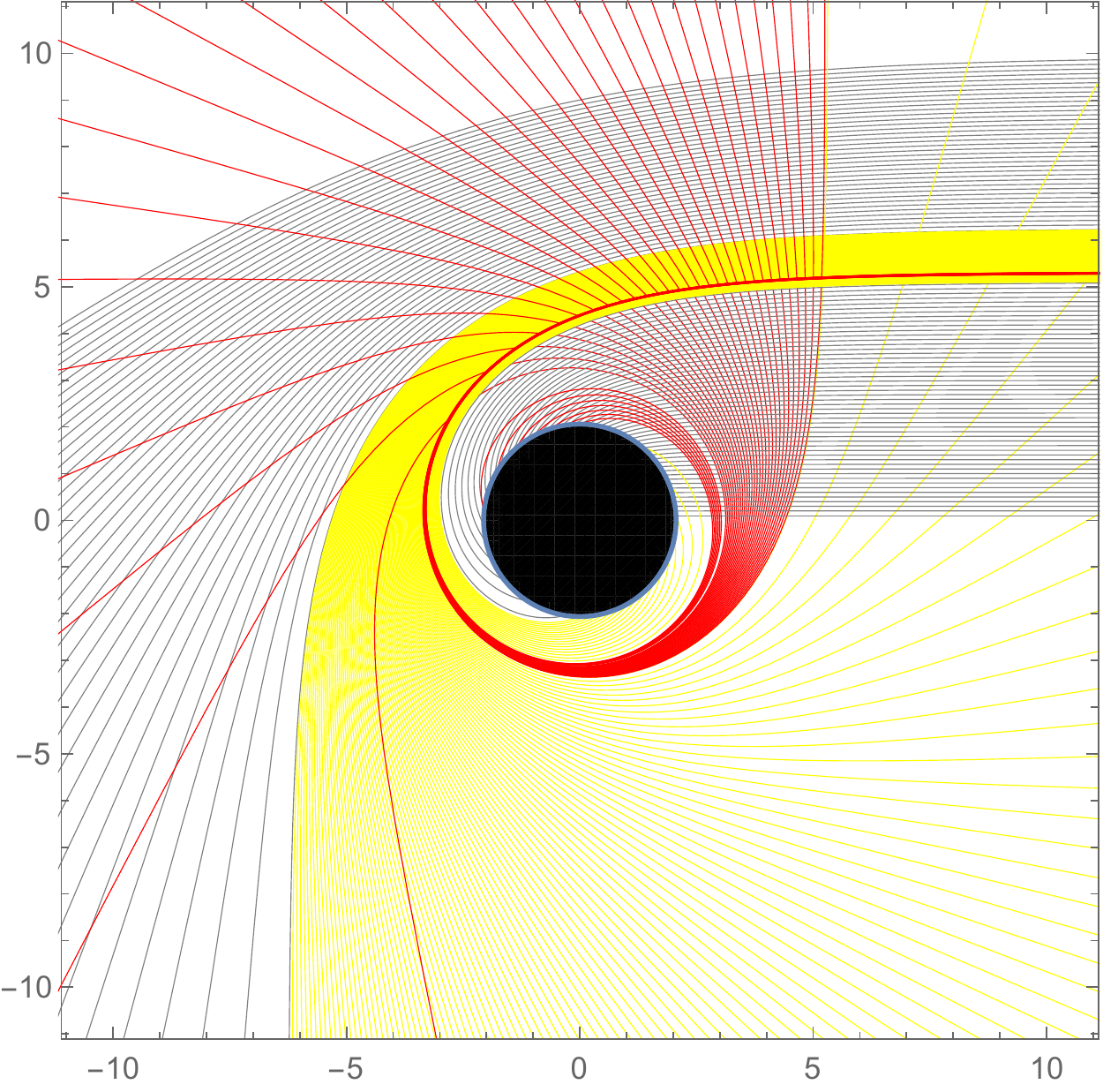}
\caption{Behavior of photons in Kazakov-Solodukhin black hole as a function of impact parameter $b$. On the left, we plot the total number of orbits $n=\phi/2\pi$. The colors correspond to $n<3/4$
(black), $3/4<n<5/4$ (yellow), and $n>5/4$ (red), defined as the direct, lensed, and photon ring trajectories, respectively. On the right we show a selection of associated photon trajectories in the Euclidean polar coordinates $(r,\phi)$. The spacing in impact parameter is 1/10, 1/100, and 1/1000 in the direct (black), lensed (yellow), and photon ring (red) bands, respectively. The black hole is
shown as a solid disk. We have set $M=1, a=0.5$.}\label{trajectory}
\end{center}
\end{figure}
In this section, to have a fuller understanding of the appearance of Kazakov-Solodukhin black hole, we move to study the black hole shadow, photon rings and lensing rings of the KS black hole around an accretion disk which has very high brightness. Hence, let's start to investigate the trajectory of a light ray traveling around the black hole. It is convenient to make a transformation $u=1/r$. The orbit equation now becomes
\be
\Big(\frac{du}{d\phi}\Big)^2=G(u)\label{geodesic}
\ee
where
\be
G(u)=\frac{1}{b^2}+2Mu^3-u^2\sqrt{1-a^2u^2}
\ee
When impact parameter $b>b_c$, the light ray move toward the black hole from infinity approaching one closest point, and then move away from the black hole back to infinity. When impact parameter $b<b_c$, the light ray always falls into the black hole. When $b=b_c$, the light ray will revolve around the black hole at radius $r_c$, i.e. the photon sphere.

For $b>b_c$, the turning point corresponds to the minimally positive real root of $G(u)=0$, which we will denote by $u_m$. According to \eqref{geodesic}, the total change of azimuthal angle $\phi$ for certain trajectory with impact parameter $b$ can be calculated by
\be
\phi=2\int_0^{u_m}\frac{du}{\sqrt{G(u)}},\quad b>b_c
\ee
For $b<b_c$, we only focus on the trajectory outside the horizon, so the total change of azimuthal angle $\phi$ is obtained by
\be
\phi=\int_0^{u_0}\frac{du}{\sqrt{G(u)}},\quad b<b_c
\ee
where $u_0=1/r_0$.

In order to discuss the observational appearance of emission originating near a black hole,  The authorsr in \cite{Gralla:2019xty} divide trajectories into direct, lensed and photon ring ones. Now we give a brief introduction. One can define the total number of orbits $n=\frac{\phi}{2\pi}$ which is obviously a function of impact parameter $b$. We denote the solution of
\be
n(b)=\fft{2m-1}{4},\quad m=1,2,3,\cdots
\ee
by $b_m^\pm$. Note that $b_m^-<b_c$ and $b_m^+>b_c$. Then we can classify all trajectories as follows:
\begin{itemize}
  \item direct: $\ft14<n<\ft34 \Rightarrow b\in(b_1^-,b_2^-)\cup(b_2^+,\infty)$
  \item lensed: $\ft34<n<\ft54 \Rightarrow b\in(b_2^-,b_3^-)\cup(b_3^+,b_2^+)$
  \item photon ring: $n>\ft54 \Rightarrow b\in(b_3^-,b_3^+)$
\end{itemize}
The physical picture of this classification is clear from the trajectory plots in Fig.\ref{trajectory}. Assuming light rays emit from north pole direction (far right of the trajectory plots), trajectories whose number of orbits $1/4<n<3/4$ will intersect the equatorial plane only once. Trajectories whose number of orbits $3/4<n<5/4$ will intersect the equatorial plane twice. Trajectories whose number of orbits $n>5/4$ will intersect the equatorial plane at least 3 times.

\section{Time-like geodesic and innermost stable circular orbit}
For time-like geodesic ${\cal L}=-1/2$, the orbit equation is
\be
\Big(\fft{dr}{d\phi}\Big)^2=V_{eff}
\ee
with the effective potential given by
\be
V_{eff}=r^4\Big(\fft{E^2}{L^2}-\fft{f(r)}{L^2}-\fft{f(r)}{r^2}\Big)
\ee
The innermost stable circular orbit (ISCO) is determined by
\be
V_{eff}=0,\quad V'_{eff}=0,\quad V''_{eff}=0
\ee
We have no an analytic expression for $r_{isco}$, but we can numerically show that $r_{isco}>6M$. Here we give a table of various important physical quantities for different deformation parameter $a$.
\begin{table}[h!]
\begin{center}
\caption{Various involved physical quantities for different deformation parameter $a$ with $M=1$}
\begin{tabular}{cccccccccc}
  \hline
  $a$ & $r_0$ & $r_{ph}$ & $r_{isco}$ & $b_c$ & $b_1^-$ & $b_2^-$ & $b_2^+$ & $b_3^-$ & $b_3^+$ \\\hline
  0 & 2 & 3 & 6 & 5.19615 & 2.8477 & 5.01514 & 6.16757 & 5.18781 & 5.22794 \\
  0.1 & 2.0025 & 3.00333 & 6.0075 & 5.20048 & 2.85105 & 5.01969 & 6.1713 & 5.19216 & 5.2322 \\
  0.5 & 2.06155 & 3.08193 & 6.18467 & 5.30259 & 2.93025 & 5.12696 & 6.25978 & 5.29484 & 5.33276 \\
  1 & 2.23607 & 3.31284 & 6.7086 & 5.60259 & 3.16347 & 5.44062 & 6.5239 & 5.59624 & 5.62886 \\
  \hline
\end{tabular}
\end{center}
\end{table}

\section{Observational appearance of thin disk emission}
Usually the term "shadow" describes the appearance of a black hole illuminated from all directions, so the shadow radius is given by critical impact parameter $b_c$. In practice, the emission is always accumulated in certain finite region near the black hole such as accretion disk. Armed with the previous preparations, now we are ready to consider a concrete accretion disk around the Kazakov-Solodukhin black hole on the equatorial plane. There are many models in accretion disk theory, however, we focus on the optically and geometrically thin disks for simplicity.

A static observer is assumed to locate at the north pole, and the lights emitted from the accretion disk is considered isotropic in the rest frame of the static observer. In view of the spherical symmetry of the spacetime, we also suppose the emitted specific intensity only depends on the radial coordinate, denoted by $I^{\rm em}_\nu(r)$ with emission frequency $\nu$ in a static frame. An observer in infinity will receive the specific intensity $I^{\rm obs}_{\nu'}$ with redshifted frequency $\nu'=\sqrt{f}\nu$. Considering $I_\nu/\nu^3$ is conserved along a ray, i.e.
\be
\frac{I^{\rm obs}_{\nu'}}{\nu'^3}=\frac{I^{\rm em}_\nu}{\nu^3}
\ee
we have the observed specific intensity
\be
I^{\rm obs}_{\nu'}=f^{3/2}(r)I^{\rm em}_\nu(r)
\ee
So the total observed intensity is an integral over all frequencies
\be
I^{\rm obs}=\int I^{\rm obs}_{\nu'}d\nu'=\int f^2 I^{\rm em}_\nu d\nu=f^2(r)I^{\rm em}(r)
\ee
where $I^{\rm em}=\int I^{\rm em}_\nu d\nu$ is the total emitted intensity from the accretion disk.

In addition, the intensity of the lights emitted from the accretion disk is so big that other sources in the environment can be ignored. If a light ray from the observer intersects with the emission disk, it means the intersecting point as a light source will contribute brightness to the observer. As discussed in section \ref{bending}, a light ray whose number of orbits $n>1/4$ will intersect with the disk on the front side. If $n$ goes larger than $3/4$, the light ray will bend around the black hole, intersecting with the disk for the second time on the back side. Further, when $n>5/4$, the light ray will intersect with the disk for the third time on the front side again, and so on. Hence, the observed intensity is a sum of the intensities from each intersection,
\be
I^{\rm obs}(b)=\sum_m f^2I^{\rm em}|_{r=r_m(b)}
\ee
where $r_m(b)$ is the so called transfer function which denotes the radial position of the $m$-th intersection with the emission disk. What needs to be emphasized is that we do not consider the absorption and reflection of light by accretion disks and the loss of light intensity in the environment in our process which is just an ideal model.

We denote the solution of orbit equation \eqref{geodesic} by $u(\phi,b)$. We will focus on the first three transfer functions which can be obtained by
\bea
r_1(b)&=&\frac{1}{u(\ft\pi2,b)},\quad b\in(b_1^-,\infty)\nn\\
r_2(b)&=&\frac{1}{u(\ft{3\pi}2,b)},\quad b\in(b_2^-,b_2^+)\nn\\
r_3(b)&=&\frac{1}{u(\ft{5\pi}2,b)},\quad b\in(b_3^-,b_3^+)
\eea
\begin{figure}[t!]
\begin{center}
\includegraphics[scale=0.5]{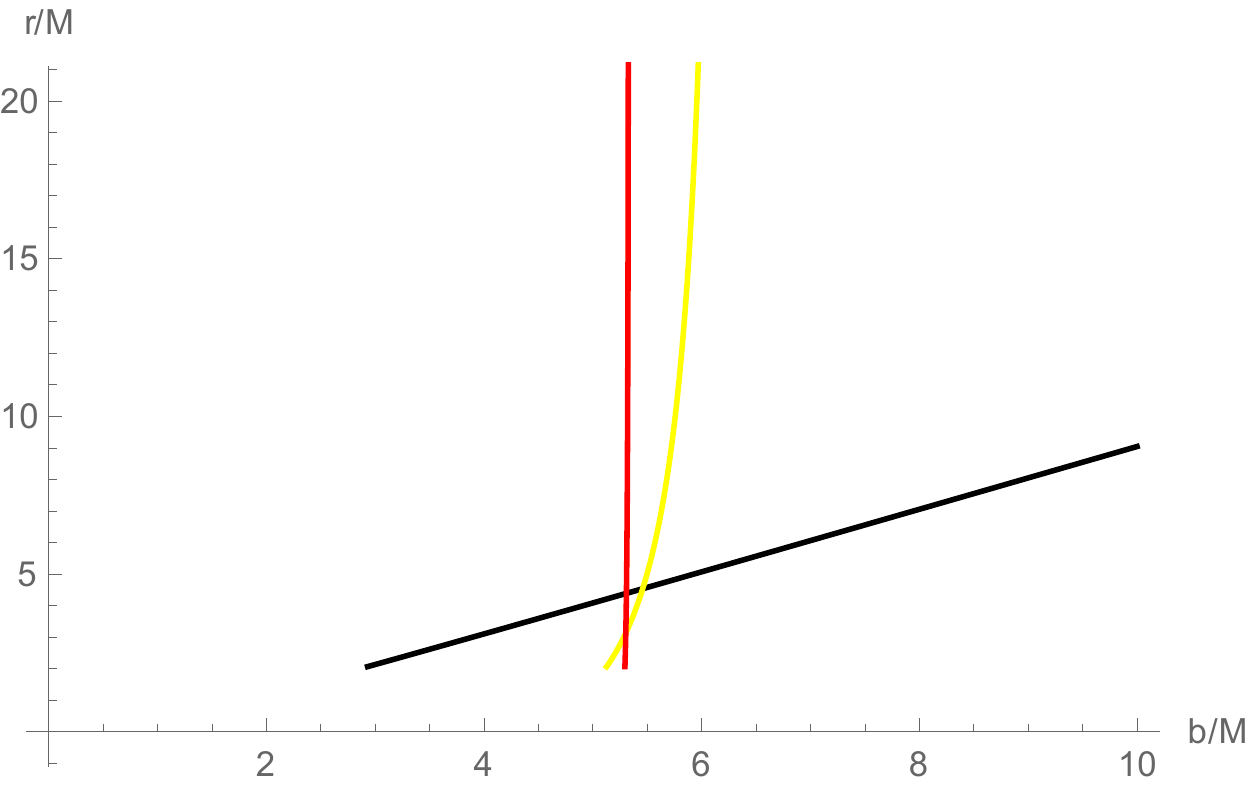}
\caption{The first three transfer functions in Kazakov-Solodukhin black hole with $M=1, a=0.5$. They represent the radial coordinate of the first (black), second (yeollow), and third (red) intersections with the emission disk.}
\end{center}
\end{figure}
As illustrated in \cite{Gralla:2019xty}, the first transfer function gives the "direct image" of the disk which is essentially just the redshift of source profile. The second transfer function gives a highly demagnified image of the back side of the disk, referred to "lensing ring". The third transfer function gives an extremely demagnified image of the front side of the disk, referred to "photon ring". The images resulted from further transfer functions are so demagnified that they can be neglected. The demagnified scale is determined by the slope of transfer function, $dr/d\phi$, called the demagnification factor.

Having obtained the transfer function, we can consider concrete emission profile. Firstly, we consider the emission is sharply peaked at the innermost stable circular orbit, and it ends abruptly at $r=r_{isco}$, such as
\be
I^{\rm em}_1(r)=\left\{
  \begin{array}{ll}
    I^0e^{-(r-r_{isco})}, & r>r_{isco} \\
    0, & r<r_{isco}
  \end{array}
\right.
\ee
Secondly, we consider the emission is sharply peaked at the photon sphere, and it ends abruptly at $r=r_c$ while decays fast to zero at $r=r_{isco}$, such as
\be
I^{\rm em}_2(r)=\left\{
  \begin{array}{ll}
    I^0\ft{2-\tanh(r-r_c)}{2}e^{-(r-r_c)}, & r>r_c \\
    0, & r<r_c
  \end{array}
\right.
\ee
Finally, we consider a emission decaying gradually from the horizon to the ISCO, such as
\be
I^{\rm em}_3(r)=\left\{
  \begin{array}{ll}
    I^0\ft{-\tanh(r-4.5)+1}{-\tanh(r_0-4.5)+1}, & r>r_0 \\
    0, & r<r_0
  \end{array}
\right.
\ee
\begin{figure}[t!]
\begin{center}
\includegraphics[scale=0.4]{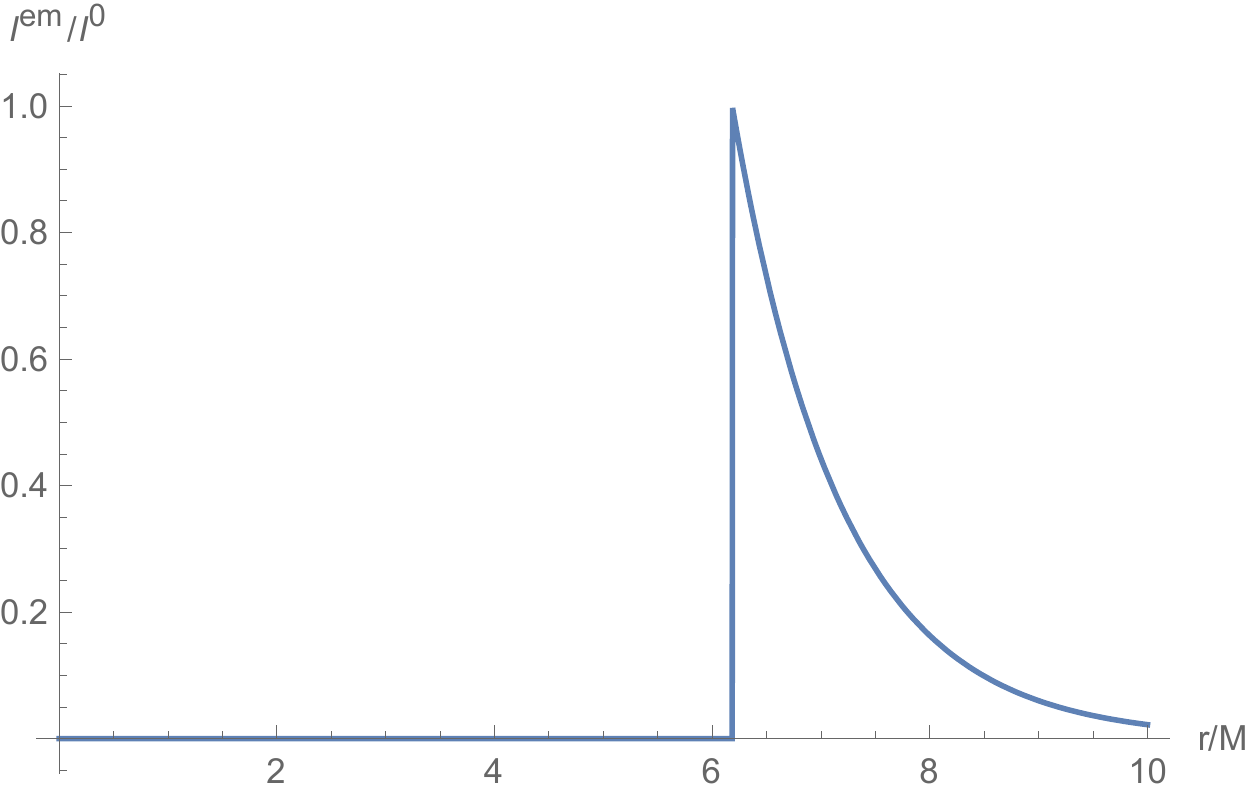}\quad\includegraphics[scale=0.4]{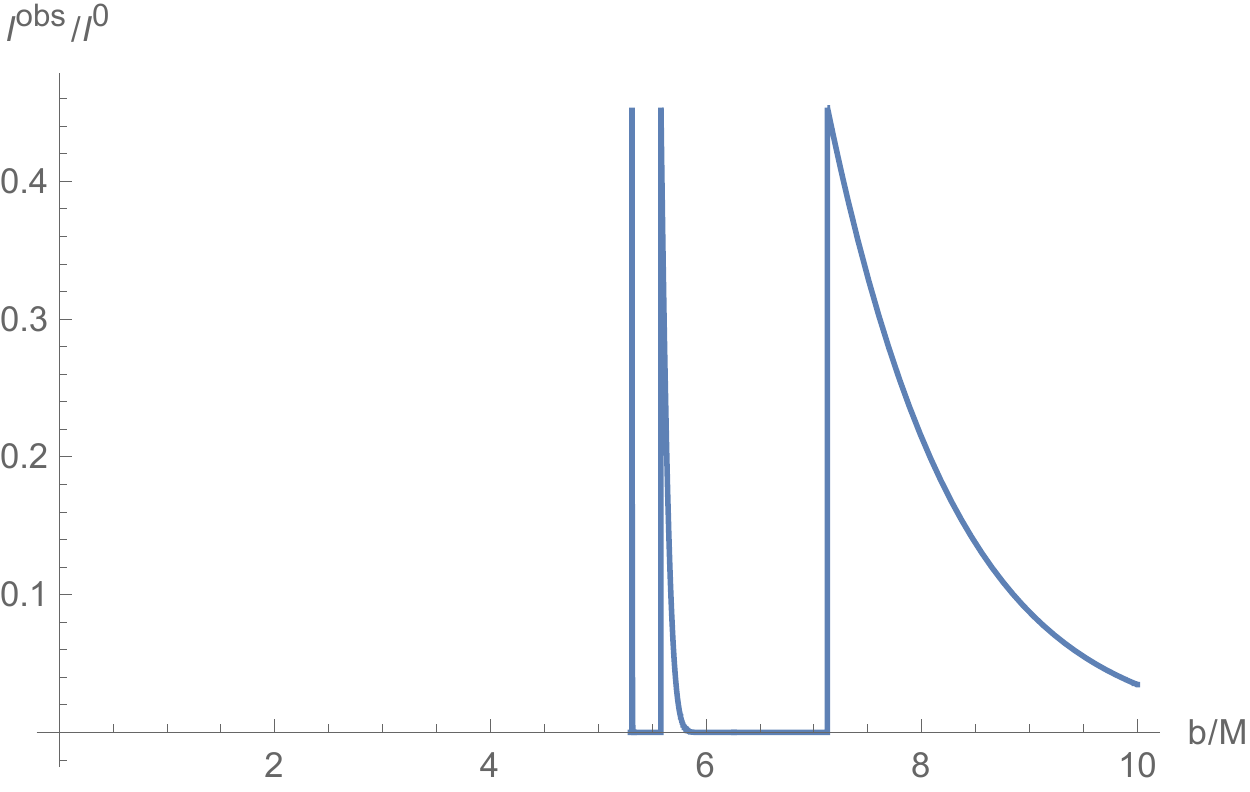}\quad\includegraphics[scale=0.3]{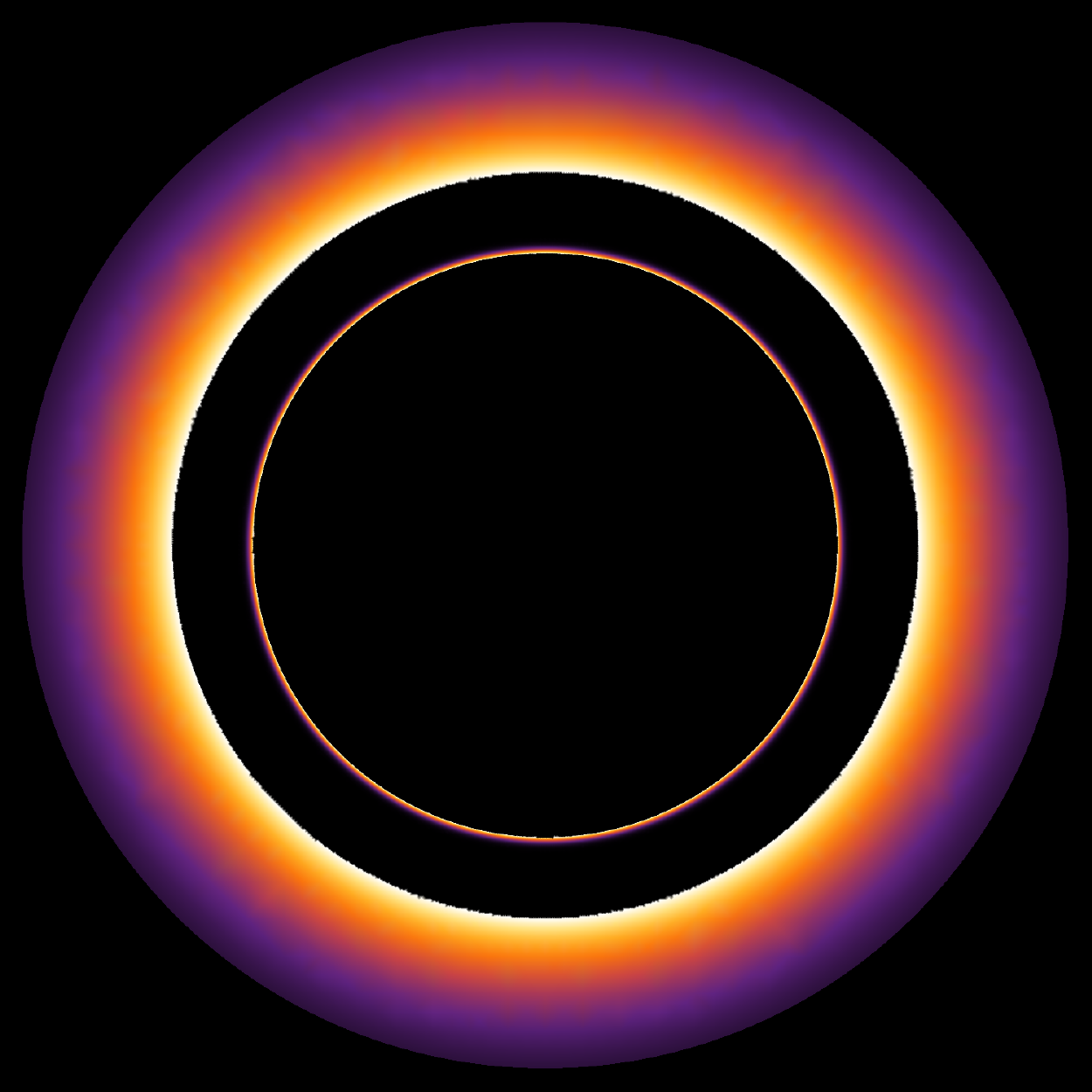}\\
\includegraphics[scale=0.4]{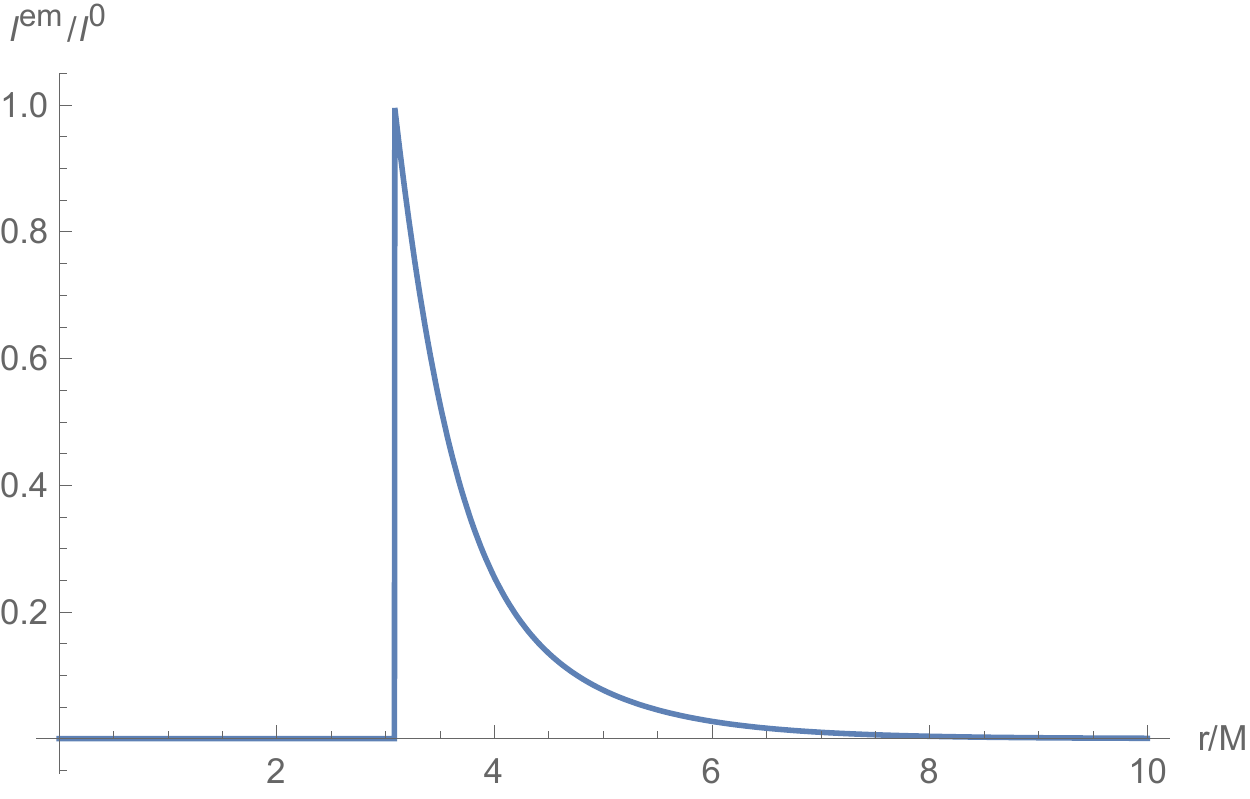}\quad\includegraphics[scale=0.4]{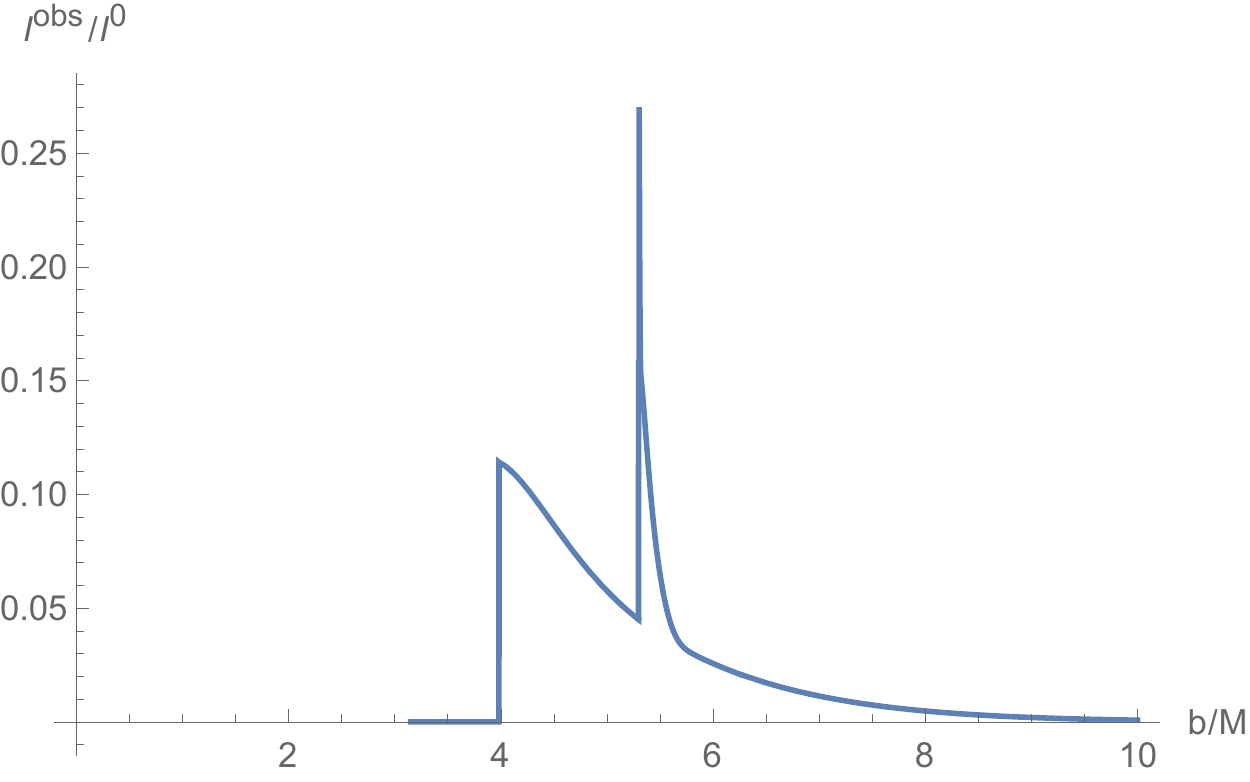}\quad\includegraphics[scale=0.3]{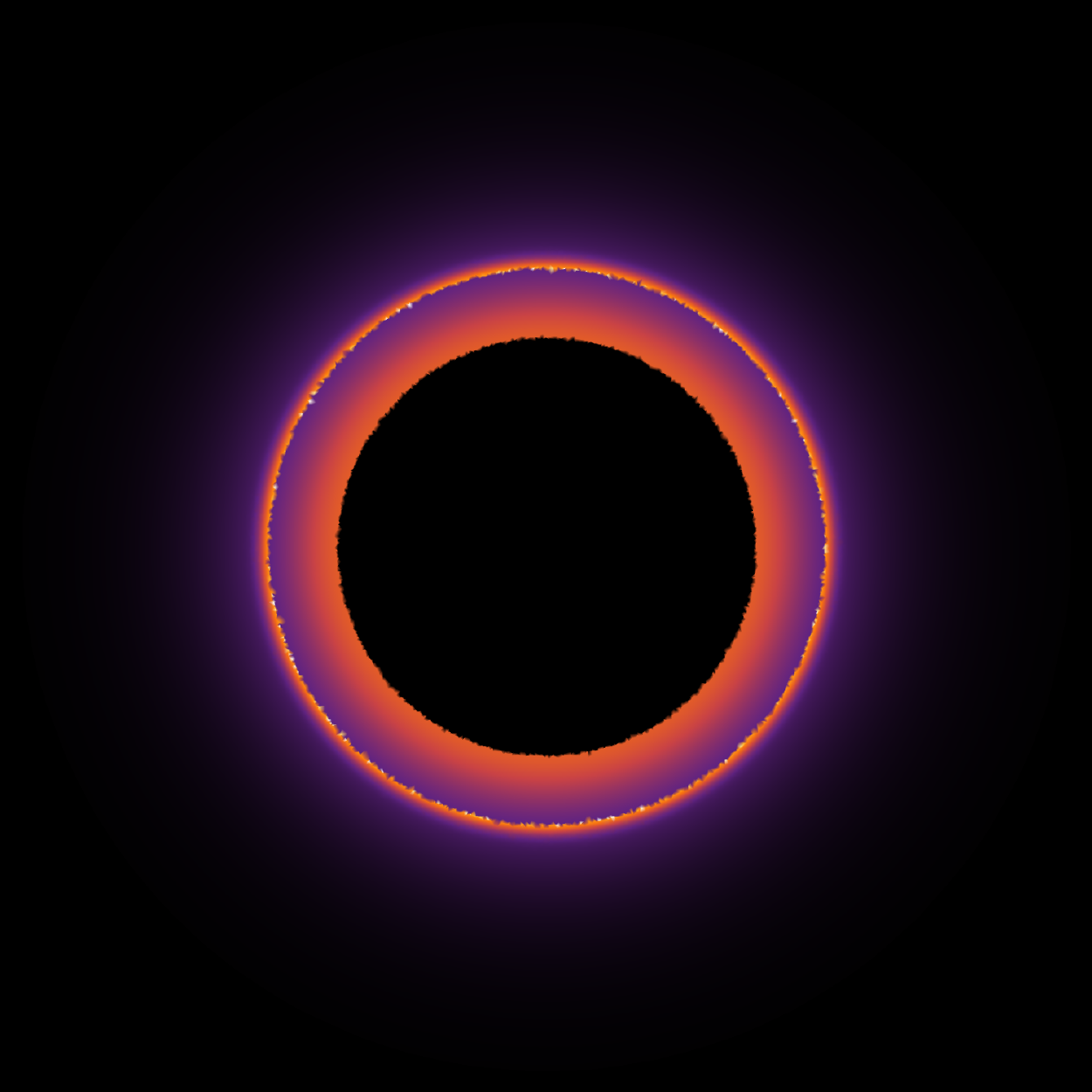}\\
\includegraphics[scale=0.4]{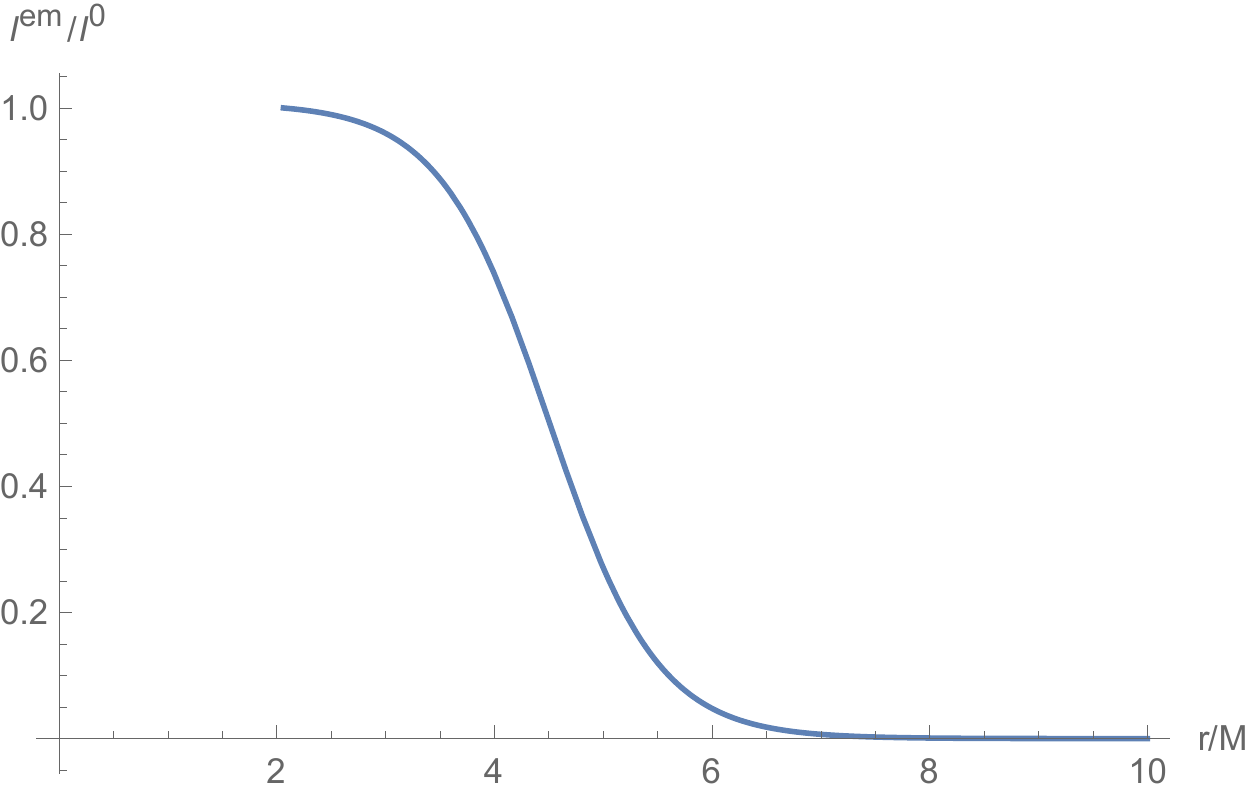}\quad\includegraphics[scale=0.4]{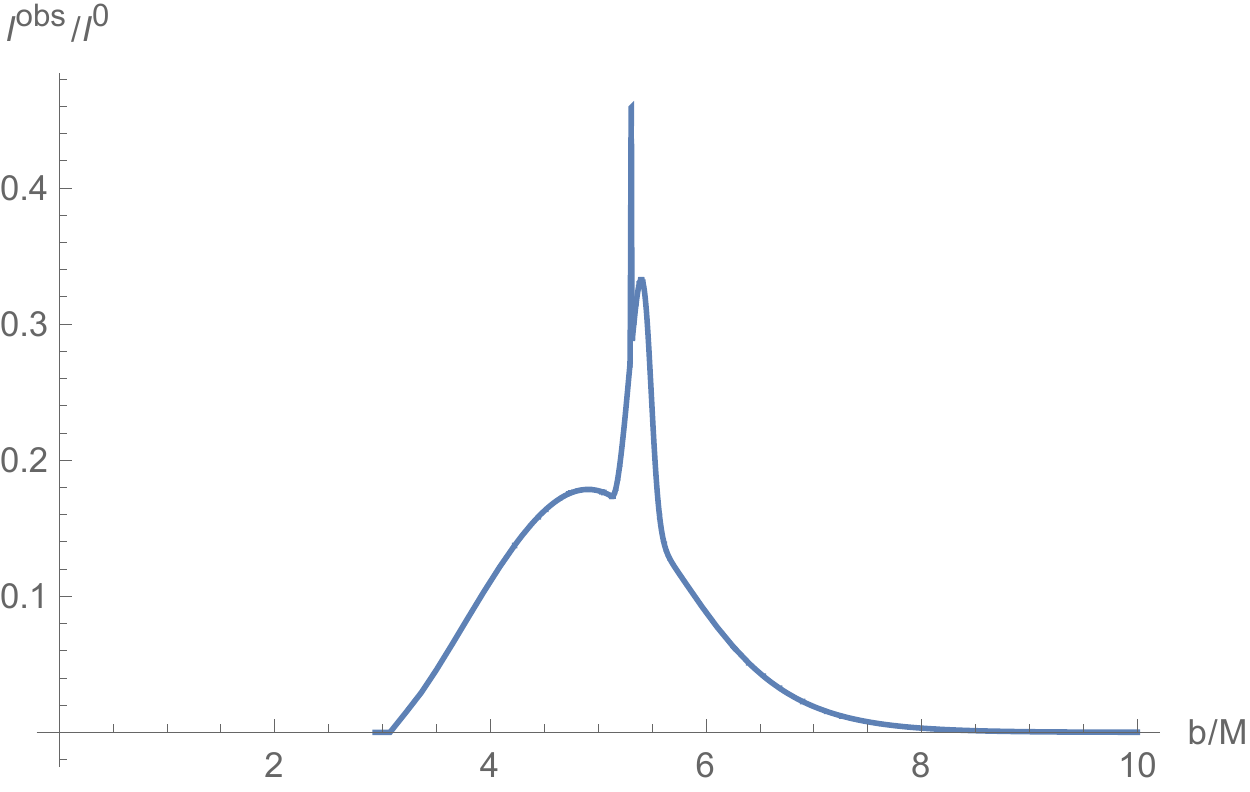}\quad\includegraphics[scale=0.3]{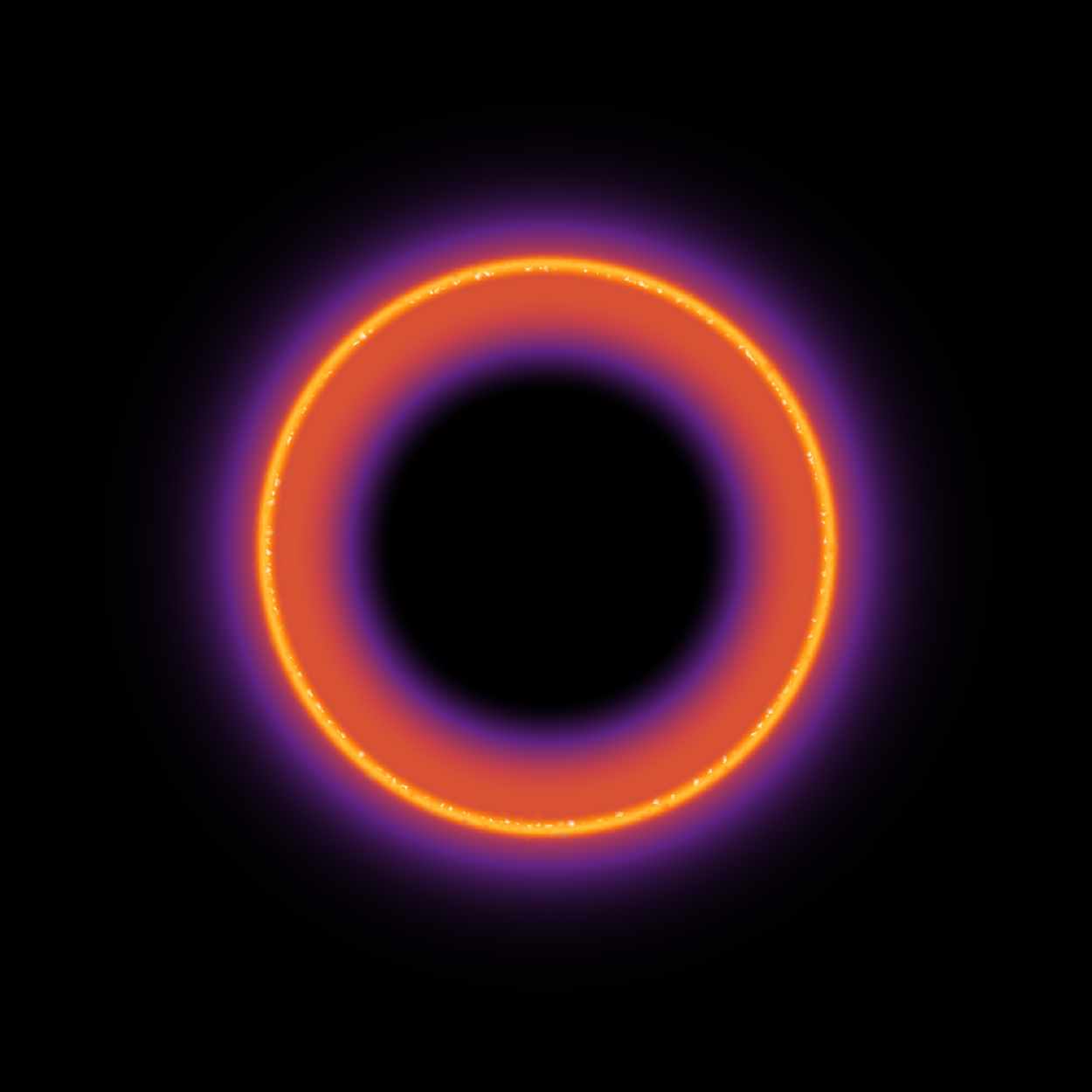}
\caption{Observational appearance of a geometrically and optically thin disk with different emission profiles near Kazakov-Solodukhin black hole with $M=1, a=0.5$. The left column are the plots of various emission profiles $I^{\rm em}(r)$. The middle column are the corresponding observed intensities $I^{\rm obs}$ as a function of impact parameter $b$. The right column are the density plots of observed intensities $I^{\rm obs}(b)$.}\label{intensity}
\end{center}
\end{figure}
We get the observed intensities $I^{\rm obs}(b)$ of these emission profiles with $M=1, a=0.5$, and plot the results in Fig.\ref{intensity}. These plots display similar features as Schwarzschild black hole. The observed intensities is dominated by the direct emission, with the lensing ring emission providing only a small contribution to the total flux and the photon ring providing a negligible contribution in all cases. The radius of the main dark area which is now called the shadow radius, is the apparent position of the edge of emission profile due to direct emission. Obviously this shadow radius is dependent on emission model. However the photon ring always occurs almost at $b_c$ and the lensing ring always occurs near $b_c$. Although the photon ring has highly enhanced brightness, it's hardly visible in the observational appearance because it is extremely demagnified. To summarise, the significant feature of observational appearance of thin disk emission near the black hole is that there exists a bright lensing ring near radius $b_c$, which is a highly demagnified image of the back of the disk.

\section{Conclusion}
The quantum correction of Schwarzschild black hole mainly affect the near horizon region of the metric, so it is very hard to detect. One important feature of the near horizon region is the strong gravitational lensing effect. There exists one critical curve with impact parameter $b_c$ which form a photon sphere with radius $r_{ph}$, and $b_c$ gives the standard shadow radius. In this paper, we find that due to quantum correction, $r_{ph}, b_c$ violate the universal inequalities \eqref{inequality} for asymptotically flat black hole which satisfy the null energy condition in the framework of Einstein gravity. However, in practice the shadow radius is dependent on the position and profile of light source.  Thus it is an unrealistic data to judge the existence of quantum correction. Nevertheless, the photon ring and lensing ring are highly associated with the critical curve $b_c$. Because of the reversed inequalities \eqref{rinequality}, we can see a larger bright lensing ring in the observational appearance of thin disk emission near the black hole compared with the classical Schwarzschild black hole. Our analysis may provide a observational evidence for quantum effect of general relativity.

We conclude this paper with some outlooks. Firstly, it's definitely interesting to investigate whether the conjecture proposed in \cite{Lu:2019zxb,Feng:2019zzn} is destroyed in other quantum corrected black hole models. If so, it' s extremely interesting to find a general proof. Secondly, a novel shadow was found from a symmetric thin-shell wormhole connecting two distinct Schwarzschild spacetimes recently in \cite{Wang:2020emr}, and one can similarly construct a thin-shell wormhole by cutting and pasting KS black hole, thus if the novel shadow exists is worthy of study when including the effects of the quantum corrections.

\section*{Acknowledgments}
J.P. is supported by the China Scholarship Council. M.G. is supported by NSFC (National Natural Science Foundation of China) Grant No. 11947210 and funded by China Postdoctoral Science Foundation Grant No. 2019M660278. X.H.F. is supported by NSFC Grant No. 11905157 and No. 11935009.

\end{document}